\documentclass[prd,aps,nofootinbib,showpacs,showkeys,preprintnumbers]
{revtex4}
\usepackage{graphicx,epsf,amsfonts,amssymb,amsbsy}
\newcommand{\vev}[1]{\langle#1\rangle}
\newcommand{\mat}{\left ( \begin{array}}
\newcommand{\emat}{\end{array} \right )}
\newcommand{\vect}{\left ( \begin{array}{c}}
\newcommand{\evect}{\end{array} \right )}

\preprint{HU-EP-05/64}
\begin{document}

\title{ \bf Pion condensation in electrically neutral cold matter
with finite baryon density}
\author{D.~Ebert}
%\email{debert@physik.hu-berlin.de}
\affiliation{Institut f\"ur Physik,
Humboldt-Universit\"at zu Berlin, 
12489 Berlin, Germany}
\author{K.\,G.~Klimenko}
%\email{kklim@ihep.ru}
\affiliation{Institute
of High Energy Physics, 142281, Protvino, Moscow Region, Russia}

\begin{abstract}
The possibility of the pion condensation phenomenon in 
cold and electrically neutral dense baryonic matter
is investigated in $\beta$-equilibrium. For simplicity, the
consideration is performed in the framework of a NJL model with two
quark flavors at zero current quark mass and for rather small values
of the baryon chemical potential, where the diquark condensation
might be ignored. Two sets of model parameters are used. For the
first one, the pion condensed phase with finite baryon density 
is realized. In this phase both electrons and the pion condensate
take part in the neutralization of the quark electric charge. For the
second set of model parameters, the pion condensation is impossible
if the neutrality condition is imposed. The behaviour of meson masses
vs quark chemical potential has been studied in electrically neutral
matter.
\end{abstract}

% 11.30.Qc Spontaneous and radiative symmetry breaking
% 12.39.-x Phenomenological quark models
% 11.15.Ex Spontaneous breaking of gauge symmetries
% 12.38.Aw General properties of QCD
% 12.38.Mh Quark-gluon plasma
% 11.10.St Bound and unstable states; Bethe-Salpeter equations
% 12.38.-t Quantum chromodynamics
% 12.38.Lg Other nonperturbative calculations
% 26.60.+c Nuclear matter aspects of neutron stars
% 21.65.+f Nuclear matter
% 12.39.Fe
% 14.40.-n  Mesons??? 

\maketitle
%\draft
%\large
%\maketitle

\section{Introduction}

According to a well-known point of view
\cite{migdal1,migdal2,others}, pionic degrees of freedom and,
especially, the pion condensation phenomenon might play 
a significant role in the description of different nuclear matter
effects. At present time it is widely believed that the dense
baryonic matter that might exist inside compact star cores or
observed in relativistic heavy ion collisions is no more than dense
quark matter, obeying an isospin asymmetry. The physics of
such a quark matter is adequately describes in the framework of QCD
with nonzero isospin chemical potential $\mu_I$. Recently, it was
shown that a nonzero pion condensate is generated in QCD if $\mu_I$
is greater than the pion mass. This result was obtained in the
framework of an effective chiral Lagrangians approach \cite{son} as
well as in QCD lattice calculations, performed at zero or small
values of the baryon chemical potential $\mu_B$ \cite{kogut}.
However, these two approaches are not applicable for the description
of an isotopically asymmetric matter at moderate baryon density. To
overcome the problem, it was proposed to study the QCD phase diagram
on the basis of Nambu -- Jona-Lasinio (NJL)-type models
\cite{njl,volkov} (see also the reviews \cite{volk,hatsuda}), which
contain quarks as microscopic degrees of freedom, in the presence of
a baryon chemical potential $\mu_B$ and an isospin $\mu_I$ one. In
this way the influence of $\mu_B$, $\mu_I$ on both the chiral
symmetry restoration effect \cite{toublan} and the formation of
color superconducting (CSC) dense baryonic matter \cite{toki} was
considered, but without taking into account the pion condensation
phenomenon. 

Recently, the pionic condensation effect was investigated in some
NJL models at nonzero values of $\mu_B$ and $\mu_I$
\cite{barducci,zhuang,ebklim}. In particular, it was shown in
\cite{ebklim} that quark matter with finite isospin density might
exist in two different phases. In the first one the baryon density
is zero and quarks are gapped, whereas in the second one the baryon
density is nonzero and quarks are gappless. Note, in
\cite{barducci,zhuang, ebklim} the chemical potentials $\mu_B$,
$\mu_I$ are independent external parameters, so the results might be
relevant to the physics of the heavy ion collision experiments only,
and do not describe the real situation inside compact stars. The
reason is that matter in the bulk of a compact star should be
electrically neutral (at least, on average) as well as remain in
$\beta$-equilibrium, i.e. all $\beta$-processes that include
quarks and leptons should go with equal rates in both directions (as
a rule, in this case $\mu_I$ depends on $\mu_B$). 

In the present paper, we study in the framework of an NJL model,
in contrast to \cite{barducci,zhuang,ebklim}, the possibility of the
pion condensation phenomenon in electrically neutral matter with
finite baryonic density at zero temperature. Moreover, matter in our
consideration is required to be in $\beta$-equilibrium. This means
that, apart from quark- and meson degrees of freedom, it is necessary
to take into account charged leptons (electrons only, for
simplicity). Since both the pion condensate and electrons have a
nonzero electric charge, it is clear that the positive charge of
quark matter might be compensated in our case by several ways,
depending on the competition between electrons and the pion
condensate. We placed the charge neutrality condition only locally,
i.e. suppose that the ground state
of matter is a uniform phase with zero electric charge density. One
should also note that we do not take into consideration the CSC
effects (see, e.g., \cite{1,hs,abuki,2,3}), so our results are valid
for not so large values of the baryon chemical potential, say, for
$\mu_B<1200$ MeV.
\footnote{The properties of the electrically neutral and
$\beta$-equilibrated CSC matter at finite baryon density 
were investigated in the framework of an NJL model in \cite{hs}, but
without taking into account the pion condensation.}

The paper is organized as follows. In Section II the phase structure
of the NJL model with zero current quark mass is investigated under
the requirements of electrical neutrality and $\beta$-equilibrium 
for the two parameter sets: $G=5.01$ GeV$^{-2}$, $\Lambda =0.65$ GeV
(set 1) and $G=6.82$ GeV$^{-2}$, $\Lambda =0.6$ GeV (set 2) ($G$ is
the model coupling constant, $\Lambda$ is the three-dimensional
cutoff parameter, used in loop integrations). It turns out that for
the set 1 the neutral matter with pion condensate is allowed to
exist at some values of $\mu\equiv\mu_B/3$, whereas for the set 2 it
is forbidden. In Section III the mass behaviours of the scalar- and
pseudoscalar mesons are considered vs $\mu$ for the parameter set 1.

\section{The model and its phase structure}

Our investigations are based on the NJL model with two
quark flavors. The corresponding Lagrangian has the following form
\begin{eqnarray}
&&  L_q=\bar q\gamma^\nu i\partial_\nu q+ G\Big [(\bar qq)^2+
(\bar qi\gamma^5\vec\tau q)^2\Big ],
  \label{1}
\end{eqnarray}
where $\tau_i$ ($i=1,2,3$) are Pauli matrices and, for simplicity,
current quark masses are taken zero. Clearly, the Lagrangian $L_q$
is invariant under transformations of the color $\rm SU_c(3)$ and
baryon $\rm U_B(1)$ groups as well as under the parity transformation
P. In addition, this Lagrangian is symmetric with respect to the
chiral $\rm SU(2)_L\times SU(2)_R$ group (chiral transformations act
on the flavor indices of quark fields only). In particular, it is
invariant under the isotopic $\rm SU(2)_I$ group as well. Moreover,
since $Q=I_3+B/2$ (in the flavor space $I_3=\tau_3/2$ is the
generator of the third isospin component, $Q={\rm diag}(2/3,-1/3)$ 
is the generator of the electric charge, and $B={\rm diag}(1/3,1/3)$
is the baryon charge generator), the electric charge is conserved too
in the NJL model (\ref{1}).

Due to the $\beta$-equilibrium requirement, we must incorporate
electrons in our consideration. So the full Lagrangian of the system
looks like
\begin{eqnarray}
&&  \bar L=L_q+\bar e\gamma^\nu i\partial_\nu e,
  \label{2}
\end{eqnarray}
where $e$ is the electron spinor field. (We suppose that electrons
are free massless particles, for simplicity.) Clearly, the Lagrangian
(\ref{2}) is well-suited for the description of different processes
in the vacuum. To study the properties of matter with nonzero baryon-
as well as electric charges, we need to modify (\ref{2}) as follows
\begin{eqnarray}
L=\bar L+\mu_BN_B+\mu_QN_Q,
  \label{3}
\end{eqnarray}
where $N_B$, $N_Q$ are baryon- and electric-charge
density operators, correspondingly, and $\mu_B$, $\mu_Q$ are
their chemical potentials. 
\footnote{The Lagrangian (\ref{3}) can be identically transformed in
the following way: $L=\bar L+(\mu_B/3+\mu_Q/6)\bar q \gamma^0q+
\mu_Q\bar qI_3\gamma^0q-\mu_Q\bar e\gamma^0e$, where $I_3$ is
presented after (\ref{1}). It is clear from this relation that
$\mu_Q$ is just the isospin chemical potential $\mu_I$.}
Evidently, it holds
\begin{eqnarray}
N_B=\bar q B\gamma^0q,~~~~~N_Q=\bar q Q\gamma^0q-\bar e\gamma^0e.
\label{4}
\end{eqnarray}
The $\mu_Q$-term in (\ref{3}) spoils the vacuum chiral $\rm
SU(2)_L\times SU(2)_R$ symmetry of the system. So, at $\mu_Q\ne 0$ 
the Lagrangian $L$ is invariant only under the reduced $\rm
U_{I_3L}(1)\times U_{I_3R}(1)$ chiral symmetry group, i.e. the
isotopic $\rm SU(2)_I$ symmetry between $u$ and $d$ quarks is absent
in medium. In this case the pion condensation phenomenon might
occur. It means, without loss of generality, that the ground state
expectation value of the form $\vev{\bar qi\gamma^5\tau_1 q}$ is
nonzero, whereas $\vev{\bar qi\gamma^5\tau_{2,3} q}= 0$ (clearly,
parity is broken in the ground state of matter with nonzero pion
condensate). Another characteristic of the ground state of dense
matter is the chiral condensate, i.e. the quantity $\vev{\bar q
q}$. When it is nonzero, the chiral symmetry is spontaneously broken
down. In the present paper, in order to establish the phase
structure of the neutral matter within the framework of the model
(\ref{3}), we restrict ourselves to the consideration of these two
condensates only.

The competition between these two condensates is governed by the
thermodynamic potential (TDP) which in the mean field approximation
has the following form (it can be obtained with ease, using, e.g.,
the technique of \cite{3}):
\begin{eqnarray}
\Omega(M,\Delta)=-\frac{\mu_Q^4}{12\pi^2}+\frac{M^2+\Delta^2}{4G}
-3\sum_a\int\frac{d^3p}
{(2\pi)^3}~|E_a|,
\label{5}
\end{eqnarray}
where the first term in the right hand side is the TDP of free
massless electrons. The summation in (\ref{5}) runs over all
quasiparticles ($a=u,d,\bar u,\bar d$), where
\begin{eqnarray}
E_u=E_\Delta^--\bar\mu,~~~~~~&& E_{\bar u}=E_\Delta^++\bar\mu,
\nonumber\\
E_d=E_\Delta^+-\bar\mu,~~~~~~&& E_{\bar d}=E_\Delta^-+\bar\mu,
\label{6}
\end{eqnarray}
and $E_\Delta^\pm=\sqrt{(E^\pm)^2+\Delta^2}$, $E^\pm=E\pm\mu_Q/2$,
$E=\sqrt{\vec p^2+M^2}$, $\bar\mu=\mu_B/3+\mu_Q/6$. The factor 3 in
front of the summation symbol in (\ref{5}) indicates the three-fold
degeneracy of each quasiparticle in color. Moreover, in order to
avoid usual ultraviolet divergences, the integration region in
(\ref{5}) is restricted by a cutoff $\Lambda$, i.e. $|\vec p|<
\Lambda$. First of all, let us fix the model parameters as follows:
$G =5.01$ GeV$^{-2}$, $\Lambda =0.65$ GeV (set 1) (Later, another
parameter set will be discussed). The gap coordinates $(M_0,
\Delta_0)$ of the global minimum point of the function $\Omega(M,
\Delta)$ are connected with condensates in the following way:
\begin{eqnarray}
M_0=-2G\vev{\bar qq},~~~~~~\Delta_0=-2Gi\vev{\bar q\gamma^5\tau_1
q}.
\label{7}
\end{eqnarray}
So if $\Delta_0$ is nonzero in the global minimum point (GMP), then
the pion condensation phase is realized. Note, the quark gap $M_0$
is just the dynamical (constituent) quark mass. From (\ref{5}) it is
possible to obtain the gap equations
\begin{eqnarray}
0=\frac{\partial\Omega (M,\Delta)}{\partial M}&\equiv&
\frac{M}{2G}-6M\int\frac{d^3p}{(2\pi)^3E}\Big\{\frac{
\theta(E_\Delta^+-\bar\mu)E^+}{E_\Delta^+}+
\frac{\theta(E_\Delta^--\bar\mu)E^-}{E_\Delta^-} \Big\},\nonumber\\
0=\frac{\partial\Omega (M,\Delta)}{\partial\Delta}&\equiv&
\frac{\Delta}{2G}-6\Delta\int\frac{d^3p}{(2\pi)^3}\Big\{\frac{\theta(
E_\Delta^+-\bar\mu)}{E_\Delta^+}+
\frac{\theta(E_\Delta^--\bar\mu)}{E_\Delta^-} \Big\}.
\label{8}
\end{eqnarray}
As it was noted in the Introduction, we are going to impose the
neutrality constraint locally, i.e. we search for the ground state
of the system, in which the electric charge density $n_Q\equiv
-\partial\Omega /\partial\mu_Q$ turns identically into zero. In
other words, we study the GMP of the function $\Omega(M,\Delta)$
under the constraint
\begin{eqnarray}
0=n_Q\equiv\frac{\mu_Q^3}{3\pi^2}+\int\frac{d^3p}{(2\pi)^3}\Big\{
\theta(\bar\mu-E_\Delta^+)+\theta(\bar\mu-E_\Delta^-)+3
\theta(E_\Delta^+-\bar\mu)\frac{E^+}{E_\Delta^+}-
3\theta(E_\Delta^--\bar\mu)\frac{E^-}{E_\Delta^-} \Big\}.
\label{9}
\end{eqnarray}
It is easily seen from the gap equations (\ref{8}) that at $\mu_Q\ne
0$ the global minimum point (GMP) of the TDP (\ref{5}) might take
only one of the following three forms in the $(M,\Delta)$-space:
{\bf i)} $(0,0)$, {\bf ii)} $(M_0,0)$, and {\bf iii)} $(0,\Delta_0)$. 
(In the ground state, corresponding to the GMP of the form {\bf i)},
both the chiral- and pion condensates are zero. The solution of the
type {\bf ii)} corresponds to the matter phase in which
only the chiral condensate is generated. Finally, in the GMP of the
form {\bf iii)} the chiral condensate is zero, but the pion
condensate is nonzero.) 
\footnote{\label{f1} If $\mu_Q =0$, then the TDP
(\ref{5}) depends effectively on the single variable
$\rho\equiv\sqrt{M^2+\Delta^2}$. So, at sufficiently small values of
$\mu_B$ the global minimum of the function $\Omega(M,\Delta)$
is achieved at all points of some circle in the $(M,\Delta)$-space.
Formally, in this case there is a freedom for selecting the GMP.
However, since at zero isospin chemical potential, i.e. at $\mu_Q
=0$, parity is a conserved quantity in the strongly interacting
physics, we suppose that in this case the pion  condensate is zero,
but the chiral one is nonzero (at rather small values of $\mu_B$),
i.e. the global minimum is placed by hand in the point of the form
{\bf ii)}, and the chirally noninvariant phase is realized (the
details of the phase structure investigation of the NJL model
(\ref{3}) with zero current quark mass and at $\mu_Q=0$ are
presented, e.g., in \cite{5}).}
If the neutrality requirement (\ref{9}) is not taken into account,
then the quantities $M_0$ and $\Delta_0$ are functions of the
chemical potentials $\mu_Q$ and $\mu_B$ which are independent
quantities. In this case the model is relevant for the description
of matter with nonzero electric charge density. 
\begin{figure}
  \centering
  \includegraphics[width=8cm]{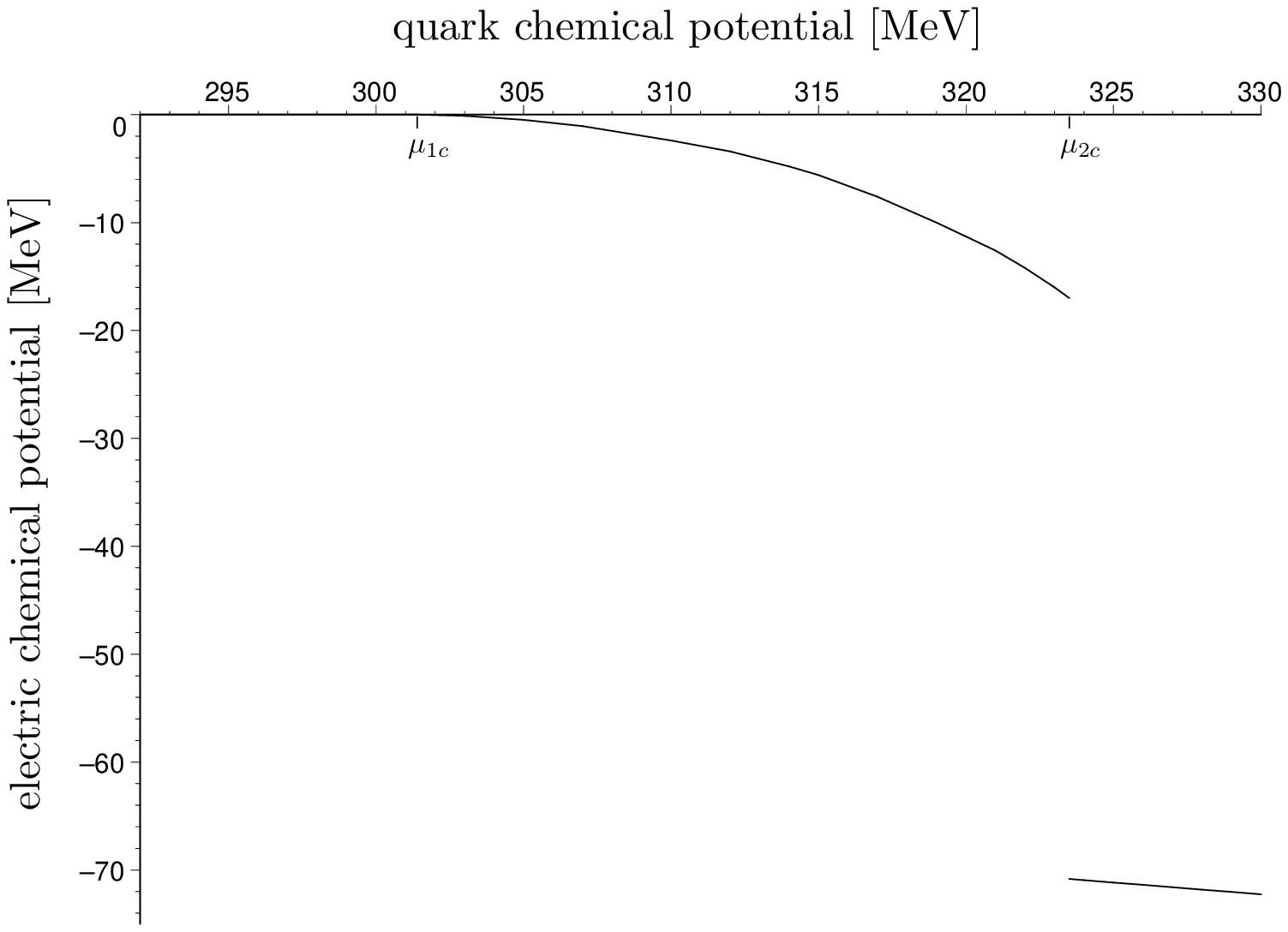}
  \caption{The behaviour of the electric chemical potential $\mu_Q$
  vs   quark chemical potential $\mu=\mu_B/3$ in the electrically
  neutral   matter for set 1 of NJL model parameters. Here
  $\mu_{1c}\approx$ 301 MeV, $\mu_{2c}\approx$ 323.5 MeV.}
\label{plot:1}
\end{figure}

\begin{figure}
  \centering
  \includegraphics[width=8cm]{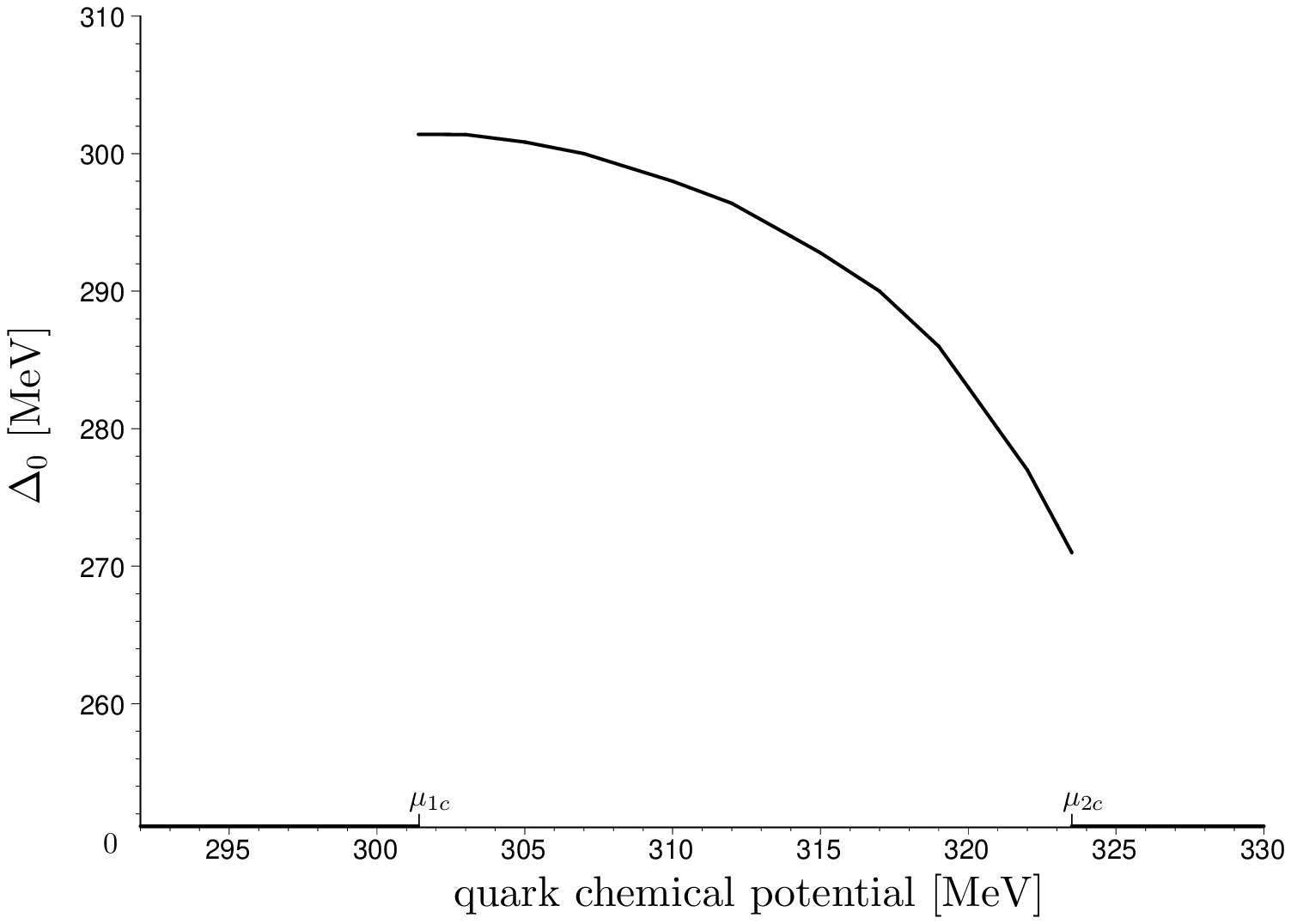}
 \caption{The pion condensate $\Delta_0$ vs $\mu=\mu_B/3$ in the
 electrically  neutral matter for the parameter set 1. Here
 $\mu_{1c}\approx$ 301  MeV, $\mu_{2c}\approx$ 323.5 MeV.}
\label{plot:2}
\end{figure}
However, since we are going to study the electrically (locally)
neutral matter, it is necessary to perform the joint consideration
of the neutrality constraint (\ref{9}) and the gap equations
(\ref{8}). In this case both the gaps, i.e. the GMP coordinates, and
the electric chemical potential $\mu_Q$ are functions of the baryon
chemical potential $\mu_B$ only (below, the analysis is performed in
terms of the quark chemical potential $\mu\equiv\mu_B/3$). Numerical
investigations of the equations (\ref{8})-(\ref{9}) show the
following results on the phase structure of the electrically neutral
matter for the parameter set 1.

At sufficiently small quark chemical potential $\mu<\mu_{1c}\approx$
301 MeV the quantity $\mu_Q$ is equal to zero (see Fig. 1). In this
case the ground state of the system corresponds to a chirally
noninvariant phase. In terms of TDP it means that at $\mu<\mu_{1c}$
the GMP has the form {\bf ii)} with $M_0\approx$ 301 MeV. In this
phase both the baryon- and the isospin densities are zeros, so there
is no need to neutralize the quark electric charge. Hence, it is not
surprising why $\mu_Q=0$ at rather small values of $\mu$. 

At $\mu_{1c}<\mu<\mu_{2c}\approx$ 323.5 MeV the GMP has already the
form {\bf iii)}, i.e. in this case the pion condensed phase is
realized. The pion condensate $\Delta_0$ vs $\mu$ is depicted in 
Fig. 2. For this phase of neutral matter the $\mu_Q$ is a
nonzero negative quantity (see Fig. 1). Nevertheless, the numerical
study shows that $\mu_Q$ is a sufficiently small quantity, so the
relation $\Delta_0<\bar\mu$ is fulfilled. As a result, one can see
that both $u$- and $d$-quasiparticles are gapless
in this phase, i.e. the quantities $E_u$ and $E_d$ from (\ref{6})
with $\Delta=\Delta_0$ and $M=0$ turn into zero at some energy
values, so there is no energy cost for creating these quasiparticles.
This fact means that at $\mu_{1c}<\mu< \mu_{2c}$ we have  the gapless
pion condensed phase (GPC) which has a nonzero baryon charge (see
also \cite{ebklim}). In the GPC phase the electric charge of quarks
is neutralized both by the electric charge of the electron gas and
the charge of the pion condensate.

Finally, at larger values of the quark chemical potential, i.e. at
$\mu_{2c}<\mu$, the normal dense baryonic phase is arranged, in
which both condensates are zero. Hence, in this phase only the
electron gas takes part in the neutralization of the quark electric
charge. Therefore, the absolute values of $\mu_Q$ at $\mu_{2c}<\mu$
are greater, than in the GPC phase (see Fig. 1). 

Now, let us consider the phase structure of the electrically
neutral matter for another set of model parameters (set 2): $G
=6.82$ GeV$^{-2}$, $\Lambda =0.6$ GeV (in this case the dynamical
quark mass is approximately 400 MeV in the vacuum, i.e. at $\mu_Q$,
$\mu_B=0$.) Numerical analysis shows that for set 2 the phase
structure of the model (\ref{3}) differs qualitatively from the set
1 case. Indeed, here at the point $\mu_c\approx 386.2$ MeV we have
at once the phase transition from the chirally noninvariant phase,
which is at $\mu<\mu_c$, to the normal dense baryonic phase, at
$\mu_c<\mu$, with zero pion- as well as chiral condensates. It turns
out that under the neutrality constraint the pion condensation is
prohibited in the NJL model with set 2 parameters at zero
temperature. 
\footnote{This is an important point. Indeed, it was shown in
\cite{hs,abuki} that in electrically neutral matter with rather large
baryon density the temperature induces a diquark condensation for
some range of model parameters, thus tending to the color
superconductivity. In a similar way, we suppose that at low baryon
density the pion condensation can appear in the neutral matter at
some temperature interval. However, it is a subject of special
consideration.}
In contrast, without this constraint the pion condensation is
allowed to exist in the set 2 NJL model (see \cite{barducci,ebklim}).

\section{Meson masses in electrically neutral matter}

In the present section the masses of mesons are investigated in
the electrically neutral matter. We will follow the way, used
in \cite{eky} for studing the particle masses in the color
superconducting quark matter. To begin with, let us
introduce auxiliary bosonic fields
\begin{eqnarray}
\sigma (x)=-2G(\bar qq),~~~\pi_a(x)=-2G(\bar qi\gamma^5\tau_a q),
\label{10}
\end{eqnarray}
where $a=1,2,3$. In the following we will ignore the influence of
electrons on the in-medium meson masses. In terms of $\sigma (x)$
and $\pi_a (x)$ the Lagrangian (\ref{3}) (with omitted electron part)
can be reduced to the form
\begin{eqnarray}
L =\bar q\Big [\gamma^\nu i\partial_\nu +\bar\mu\gamma^0
+ \frac{\mu_Q}{2}\tau_3\gamma^0-\sigma -i\gamma^5\pi_a\tau_a \Big ]q
-\frac{1}{4G}\Big [\sigma\sigma+\pi_a\pi_a\Big ]
\label{11}
\end{eqnarray}
(the quantity $\bar\mu$ is defined after (\ref{6})). Starting from
(\ref{11}), it is possible to integrate out the quark fields and
obtain the effective action of the system in the one-quark  loop
approximation:
\begin{equation}
{\cal S}_{\rm {eff}}(\sigma,\pi_a)
=-\int d^4x\left[\frac{\sigma^2+\pi^2_a}{4G}\right]-i{\rm
Tr}_{sfcx}\ln D,
\label{12}
\end{equation}
where
\begin{equation}
D=\gamma^\nu i\partial_\nu +\bar\mu\gamma^0
+ \frac{\mu_Q}{2}\tau_3\gamma^0-\sigma -i\gamma^5\pi_a\tau_a.
\label{13}
\end{equation}
The Tr-operation in (\ref{12}) stands for calculating  the trace in
spinor- ($s$), flavor- ($f$), color- ($c$) as well as
four-dimensional coordinate- ($x$) spaces, correspondingly. 

It is clear from (\ref{7}) and (\ref{10}) that the coordinates
$(M_0,\Delta_0)$ of the global minimum point of the TDP are 
just the ground state expectation values of the $\sigma$- and
$\pi_1$-fields, i.e. $M_0\equiv\vev{\sigma(x)}$, $\Delta_0\equiv
\vev{\pi_1(x)}$. 

Let us make the following field shifts in
(\ref{12}): $\sigma(x)\to M_0+\sigma(x)$, $\pi_1(x)\to \Delta_0
+\pi_1(x)$, and thereafter expand the effective action up to 
second order in the meson fields. Differentiating twicely the
obtained expression with respect to meson fields, it is then
possible to get the one-particle irreducible (1PI) Green's functions
$\Gamma_{XY}$ of the mesons ($X,Y=\sigma, \pi_1,\pi_2,\pi_3$). (In
the present paper we omit these cumbersome calculations, referring to
the similar meson mass calculations in \cite{eky}.)
The results are the following.

First, let us consider the masses of the $\sigma$- as well as
$\pi_3$-mesons. (Note, $\pi_0\equiv\pi_3$.) It turns out that both
$\sigma$- and  $\pi_0$-mesons are not mixed with other particles.
Moreover, in the momentum space representation and at zero
three-momentum, $\vec p=0$, we have in the neutral gapless pion
condensed phase of matter ($\mu_{1c}<\mu<\mu_{2c}$):
\begin{eqnarray}
&&\Gamma_{\sigma\sigma}(p_0)=\Gamma_{\pi_0\pi_0}(p_0)=\frac{1}{2G}
+6\int\frac{d^3q}{(2\pi)^3}\left\{\frac{\theta
(\varepsilon_\Delta^+-\bar\mu)}
{\varepsilon_\Delta^+} \left [\frac{\varepsilon_\Delta^+p_0-\mu_Q
\varepsilon^+}{(\varepsilon_\Delta^+-p_0)^2-(\varepsilon_\Delta^-)^2}
-\frac{\varepsilon_\Delta^+p_0+\mu_Q \varepsilon^+}
{(\varepsilon_\Delta^++p_0)^2-(\varepsilon_\Delta^-)^2}\right
]+\right. \nonumber\\&&\left.~~~~~~~~~~~+\frac{\theta
(\varepsilon_\Delta^--\bar\mu)}{\varepsilon_\Delta^-}\left [
\frac{\varepsilon_\Delta^-p_0+\mu_Q \varepsilon^-}
{(\varepsilon_\Delta^--p_0)^2-(\varepsilon_\Delta^+)^2}
+\frac{\mu_Q \varepsilon^- -\varepsilon_\Delta^-p_0}
{(\varepsilon_\Delta^-+p_0)^2-(\varepsilon_\Delta^+)^2}\right
]\right\},
\label{14}
\end{eqnarray}
where
$\varepsilon_\Delta^\pm=\sqrt{(\varepsilon^\pm)^2+\Delta_0^2}$,
$\varepsilon^\pm=|\vec q|\pm\mu_Q/2$. Recall, $\bar\mu=\mu+\mu_Q/6$.
Eliminating in (\ref{14}) the coupling constant $G$ with the help of
the $\Delta_0$-gap equation (\ref{8}), we obtain in the GPC phase
\begin{eqnarray}
&&\Gamma_{\sigma\sigma}(p_0)=\Gamma_{\pi_0\pi_0}(p_0)\sim
(p_0^2-\mu_Q^2).
\label{15}
\end{eqnarray}
Since the zero of a 1PI function in the $p_0^2$-plane defines 
the mass squared of a particle, it is evident from (\ref{15}) that
in the GPC phase $M_{\pi_0}=M_{\sigma}=|\mu_Q|$ (see Fig. 3, where
$M_{\sigma,\pi_0}$ are depicted, or Fig. 1 for $\mu_Q$).

The expressions for the 1PI functions of the $\sigma$- and $\pi_0$
mesons in the normal dense quark matter, i.e. at $\mu_{2c}<\mu$, 
follow from (\ref{14}) at $\Delta_0=0$. The zeros of the Green's
functions $\Gamma_{\sigma\sigma}(p_0)$ and $\Gamma_{\pi_0\pi_0}(p_0)$
in the $p_0^2$-plane were studied numerically in this phase as well.
The corresponding mass behaviours vs $\mu$ are depicted in Fig. 3 at
$\mu_{2c}<\mu$. 

In contrast to the $\sigma ,\pi_0$-sector, in the sector of $\pi_1$
and $\pi_2$ fields the 1PI Green's functions (at $\vec p=0$) form a
nontrivial matrix $\Gamma^{\rm GPC} (p_0)$ in the GPC phase, i.e.
there is a mixing between $\pi_1$ and $\pi_2$. \footnote{Note, at
nonzero current quark mass there is actually a mixing between
$\sigma$-, $\pi_1$- and $\pi_2$ fields in the pion condensed phase.}
Its matrix elements are:
\begin{eqnarray}
\Gamma^{\rm GPC}_{\pi_1\pi_1}(p_0)=6(p_0^2-4\Delta_0^2)A(p_0^2),~~
\Gamma^{\rm GPC}_{\pi_2\pi_2}(p_0)=6p_0^2A(p_0^2),~~
\Gamma^{\rm GPC}_{\pi_2\pi_1}(p_0)=\Gamma^{\rm GPC}
_{\pi_1\pi_2}(-p_0)=12ip_0B(p_0^2),
\label{16}
\end{eqnarray}
where $\Delta_0$ is presented in Fig. 2 and
\begin{eqnarray}
A(p_0^2)&=&\int\frac{d^3q}{(2\pi)^3}\left\{
\frac{\theta (\varepsilon_\Delta^+-\bar\mu)}{\varepsilon_\Delta^+
[p_0^2-4(\varepsilon_\Delta^+)^2]}
+\frac{\theta (\varepsilon_\Delta^--\bar\mu)}{\varepsilon_\Delta^-
[p_0^2-4(\varepsilon_\Delta^-)^2]}\right\},\nonumber\\
B(p_0^2)&=&\int\frac{d^3q}{(2\pi)^3}\left\{
\frac{\varepsilon^+\theta
(\varepsilon_\Delta^+-\bar\mu)}{\varepsilon_\Delta^+
[p_0^2-4(\varepsilon_\Delta^+)^2]}-\frac{\varepsilon^-\theta
(\varepsilon_\Delta^--\bar\mu)}{\varepsilon_\Delta^-
[p_0^2-4(\varepsilon_\Delta^-)^2]}\right\}
\label{17}
\end{eqnarray}
(see also notations after (\ref{14})). Note, in the GPC phase
$\Delta_0<\bar\mu$ and, in addition, $\mu_Q<0$. Therefore, the
minimal value of the quantity $\varepsilon_\Delta^+$ in the
integrands of (\ref{17}) is $\bar\mu$. As a result, we see that
$A(p_0^2)$ and $B(p_0^2)$ are analytical functions in the whole
$p_0^2$-plane, except the cut which is at $p_0^2>4\bar\mu^2$. Since
there is a mixing between $\pi_1$ and $\pi_2$ fields, the masses of
meson modes in this sector are defined by the zeros of the
det($\Gamma^{\rm GPC} (p_0)$) in the $p_0^2$-plane, i.e. by the
equation  
\begin{equation}
  \label{18}
{\rm det}(\Gamma^{\rm GPC} (p_0))= 36p_0^2\big\{(p_0^2-4\Delta_0^2 )
A^2(p_0^2)-4B^2(p_0^2)\big\}=0.
\end{equation}
The evident solution of this equation is $p_0^2=0$. It corresponds
to a massless meson mode, specified by $\pi_L$ (see Fig. 3), that is
actually the Nambu-Goldstone boson. (The appearence of such a mode
in the GPC phase, is justified by the spontaneous breaking of the
initial $\rm U_{I_3L}(1)\times U_{I_3R}(1)$ chiral symmetry down to
the abelian subgroup.) The nontrivial solution of (\ref{18}) is the
zero of the expression in the braces. It corresponds to a massive
meson mode, denoted by $\pi_H$ (see Fig. 3). (Clearly, its mass,
$M_{\pi_H}$, lies in the interval $2\Delta_0<M_{\pi_H}<2\bar\mu$.)

In the dense normal quark matter (NQM) phase, i.e. at at
$\mu_{2c}<\mu$, it is convinient to use the charged fields
$\pi_\pm(x)=(\pi_1(x)\pm i\pi_2(x))/\sqrt{2}$. Then in the NQM phase
the matrix $\Gamma^{\rm NQM} (p_0)$ of 1PI Green's functions of the
$\pi_\pm$ mesons looks like: 
\begin{eqnarray}
&&\Gamma^{\rm NQM}_{\pi_+\pi_-}(p_0)=\Gamma^{\rm NQM}_{\pi_-\pi_+}
(-p_0)=
\frac{1}{2G}-\frac{3}{\pi^2}\Big\{\Lambda^2-\frac{3\mu_Q^2}{4}-
\frac{p_0\mu_Q}{2}-\left (\mu+\frac{\mu_Q}{6}\right )^2+\nonumber\\
&&+\frac{(\mu_Q+p_0)^2}{4}\ln\left
[\frac{4\Lambda^2-(\mu_Q+p_0)^2}{(2\mu+\mu_Q/3)^2-p_0^2}\right
]\Big\},~~~~~\Gamma^{\rm NQM}_{\pi_+\pi_+}(p_0) = \Gamma^{\rm
NQM}_{\pi_-\pi_-}(p_0)=0.
  \label{19}
\end{eqnarray}
The numerical investigation of the zeros of the quantity
det($\Gamma^{\rm GPC} (p_0)$) shows the presence of two pionic
massive modes, which might be identified in NQM phase with
$\pi_\pm$-mesons (see Fig. 3). Evidently, in this phase the mass
splitting between $\pi_\pm$-mesons is due to the isospin asymmetry
that is generated by nonzero $\mu_Q$.

Finally, in Fig. 3 the masses of mesons in the chirally
noninvariant phase, i.e. at $\mu<\mu_{1c}$, are presented as well. 
In this phase $\mu_Q\equiv 0$, and only the chiral gap $M_0\ne 0$.
It is well-known that in this case $M_\sigma =2M_0\approx$ 602 MeV,
and the three $\pi$-mesons are massless Nambu--Goldstone bosons. 

\begin{figure}
  \centering
  \includegraphics[width=8cm]{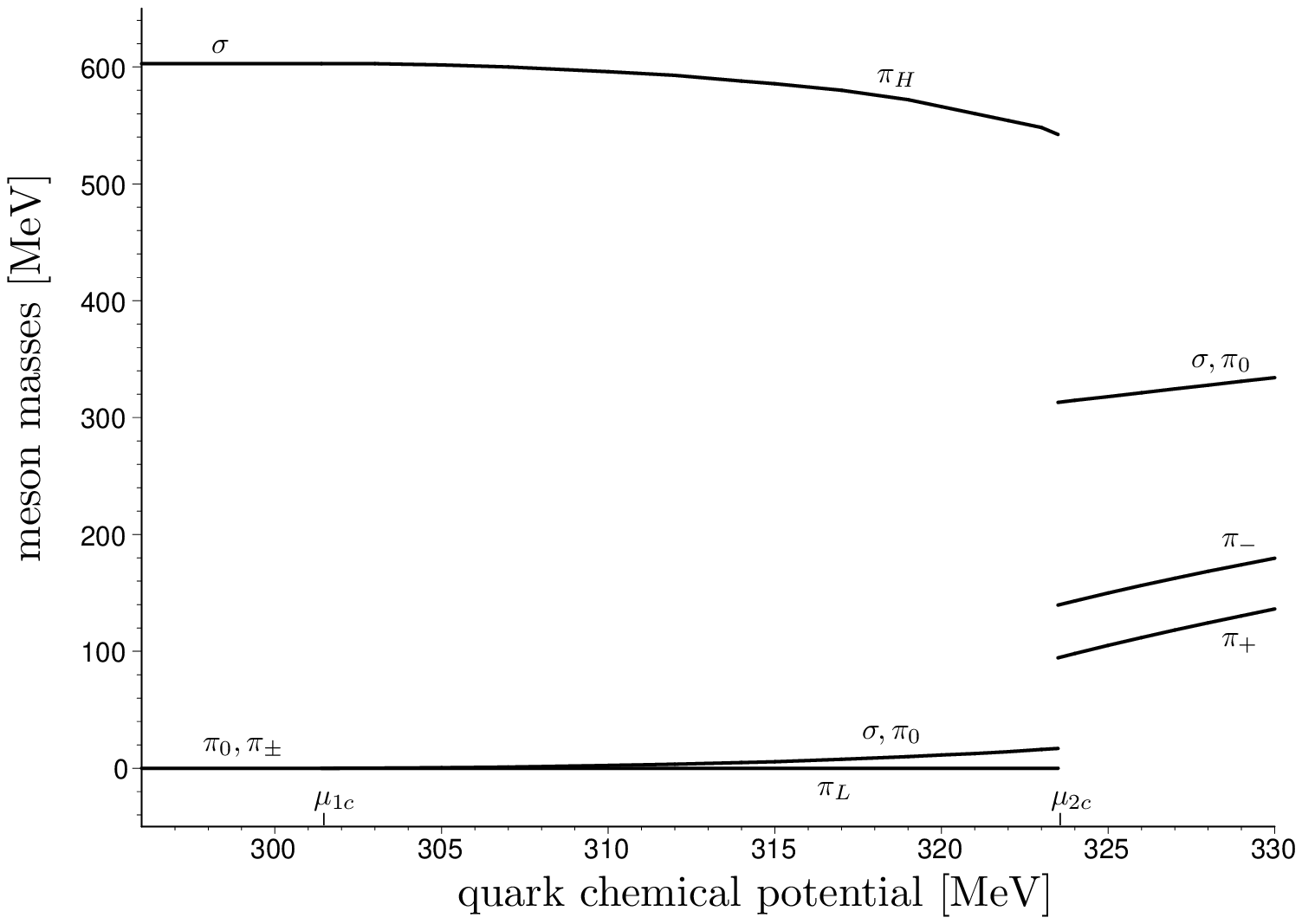}
  \caption{Scalar- and pseudoscalar meson masses vs $\mu=\mu_B/3$ in
  the electrically neutral matter for the parameter set 1. Here
  $\mu_{1c}\approx$ 301 MeV, $\mu_{2c}\approx$ 323.5 MeV.}
\label{plot:3}
\end{figure}

\section{Summary}

In the present paper we have studied the properties of electrically
neutral and $\beta$-equlibrated cold matter with finite baryonic
density. The problem is inspired by the physics of compact stars. For
simplicity, the consideration was done in the framework of a NJL
model with zero current quark mass. We have found that for the set 1
of model 
parameters (see the end of Introduction) there are three
different phases, including the one with a pion condensate, of
neutral matter. In contrast, for the parameter set 2 the pion
condensation in the neutral matter is forbidden. Moreover, we have 
studied the behavior of meson masses vs quark chemical potential in
the case of the parameter set 1 (see Fig. 3). Since the electric
neutrality of the system is realized together with an isospin
asymmetry between quarks, it turns out that the masses of
$\pi$-mesons are splitted at nonzero baryon density.

{\bf Acknowledgments:} 
This work was supported in part by DFG-project 436 RUS 113/477/0-2
and RFBR grant No. 05-02-16699.

\end{document}